\documentstyle[epsfig,11pt]{article}

\def\chpt{{$\chi PT$}}

\def\roughly#1{\mathrel{\raise.3ex\hbox{$#1$\kern-.75em%
\lower1ex\hbox{$\sim$}}}}

\def\lsim{\roughly<}

\def\be{\begin{eqnarray}}
\def\ee{\end{eqnarray}}
\def\ben{\begin{enumerate}}
\def\een{\end{enumerate}}
\def\beitem{\begin{itemize}}
\def\eitem{\end{itemize}}
\def\bitem{\begin{itemize}}
\def\eitem{\end{itemize}}

\newcommand{\beq}{\begin{eqnarray}}
\newcommand{\eeq}{\end{eqnarray}}
\def\la{\langle}
\def\ra{\rangle}
\def\bi{\begin{itemize}}
\def\ei{\end{itemize}}

\def\etal{{\it et al}}
\def\del{\partial}

\long\def\beginomit#1\endomit{}
\def\np{{Nucl. Phys.}}
\def\prl{Phys. Rev. Lett.}
\def\pr {Phys. Rev.}

\def\pl{Phys. Lett.}

\def\PR{{ Phys. Repts.}}
\def\chpt{$\chi$PT}

\def\etal{{\it et al.}}

\def\del{\partial}

\begin{document}

%-------------------------------------------------------------------------%
%                     Title Page                                          %
%-------------------------------------------------------------------------%
\begin{titlepage}\begin{center}

\hfill{T94/150}

\hfill{hep-ph/9501239}

\vskip 0.6in
{\large\bf EFFECTIVE FIELD THEORIES FOR HADRONS AND NUCLEI\footnote{
Invited talk given at {\it Nishinomiya Yukawa-Memorial Symposium in
Theoretical Physics}, October 27-28, 1994, Nishinomiya, Japan.}}
\vskip 0.8in
{\large  Mannque Rho}\\
\vskip 0.1in
{\large  \it Service de Physique Th\'{e}orique, CEA  Saclay}\\
{\large\it 91191 Gif-sur-Yvette Cedex, France}\\
\vskip .6in
\centerline{December 1994}
\vskip .6in

{\bf ABSTRACT}\\ \vskip 0.1in
\begin{quotation}
\noindent
Hadron structure and nuclear structure are discussed from the common ground
of effective chiral Lagrangians modeling QCD at low energy. The topics treated
are the chiral bag model in large $N_c$ QCD, its connection to heavy-baryon
chiral perturbation theory (HB\chpt),\ the role of nonabelian Berry gauge
connections for baryon excitations and the application of HB\chpt \ to
the thermal $n+p\rightarrow d +\gamma$ process and to the axial-charge
transitions in heavy nuclei.

\end{quotation}
\end{center}\end{titlepage}

\section{Introduction}
\indent

The quarks that enter into nuclei and hence figure in
nuclear physics are the u(p),
d(own) and possibly s(trange) quarks. These are called ``chiral quarks"
since they are very light at the scale of strong interactions. Both the u and
d quarks are less than 10 MeV, much less than the relevant scale which I will
identify with the vector meson (say, $\rho$) mass $\sim 1$ GeV. The s quark
is in the range of 130 to 180 MeV, so it is not quite light. In some sense,
it may be classified as ``heavy" as in the Skyrme model for hyperons but
the success with current algebras involving kaons also indicates that it can be
considered as chiral as the u and d are. In this talk, I will consider the
s quark as both heavy -- when I describe baryon excitations and light
-- when I describe kaon condensation.

If the quark masses are zero, the QCD Lagrangian has chiral symmetry
$SU(n_f)\times SU(n_f)$ where $n_f$ is the number of massless flavors.
However we know that this symmetry in the world we are living in, namely
at low temperature ($T$) and low density ($\rho$), is {\it spontaneously}
broken to $SU(n_f)_V$ giving rise to Goldstone bosons denoted $\pi$,
the pions for $n_f=2$, the pions and kaons for $n_f=3$.
In nature, the quark masses are not
strictly zero, so the chiral symmetry is explicitly broken by the masses,
and the bosons are pseudo-Godstones with small mass. Again in the u and d
sector, the pion is very light $\sim 140$ MeV but in the strange
sector, the kaon is not so light, $\sim 500$ MeV. Nonetheless we will
pretend that we have good chiral symmetry and rectify our mistakes
by adding symmetry breaking terms treated in a suitable way.

The theme of this talk is then that most of what happens in nuclei are
strongly controlled by this symmetry pattern. Indeed, it was argued many
years ago\cite{rhobrown81} that chiral symmetry should play a
crucial role in many nuclear processes, much more than
confinement and asymptotic freedom -- the other basic ingredients of QCD --
would. More recently, it has become clear that much of what we can understand
of the fundamental nucleon structure also follows from chiral symmetry and its
breaking. This was also anticipated sometime ago\cite{brtoday,br-comm88}.

In this talk, I would like to tell you more recent and quite exciting
new development in this line of work which suggests that the old idea, quite
vague at the beginning, is becoming a viable model of QCD in many-body nuclear
systems.
\section{Nucleon Structure: The Chiral Bag in QCD}
\indent

The chiral bag was formulated originally in a somewhat as hoc way based solely
on chiral symmetry but there is a striking indication\cite{mattis,manohar}
that it follows from a
more general argument based on large $N_c$ QCD where $N_c$ is the number of
colors. Let me discuss this as a model for nucleon structure.

When chiral symmetry is implemented to the bag model of the
hadrons\cite{br-comm88,chodos}, it was found necessary to introduce pion fields
outside of the bag of radius $R$ in which quarks are ``confined."
This is because otherwise the axial current cannot be conserved.
Furthermore, it was discovered\cite{rgb83} that to be consistent with
non-perturbative structure, the pion field takes the form of the
skyrmion configuration with a fractional baryon charge residing in the
pion cloud. This implied that the quarks are not strictly confined
in the sense of the MIT bag but various charges leak out. It became
clear that the bag radius is  not a physical variable. That physics should
not depend upon the size of the bag has been formulated as
a ``Cheshire Cat Principle" (CCP). In fact the CCP may be stated as a
gauge principle\cite{damgaardCCP} with the bag taken as a gauge fixing.
What this
meant was that the skyrmion is just a chiral bag whose radius is
``gauge-chosen" to be shrunk to a point.

\def\Nc{$N_c$}

The recent development\cite{mattis,manohar} is closer to the core of
QCD. In large $N_c$ QCD, meson-meson interactions become weak but
meson-baryon Yukawa interactions become strong going like $N_c^{1/2}$.
In this limit, the baryon is heavy and hence can be treated as a static
source localized at the origin. Other interactions, such as mass splittings
etc. are down by a factor of {$N_c$}. Thus
we have to add to the usual current algebra
Lagrangian of $O(N_c)$, $L_{ca}$,
a term of the form\cite{manohar}
\be
\delta L=3g_A \delta(\vec{x}) X^{ia} A^{ia} (x)
\ee
where $X^{ia}$ is the baryon axial current in the large $N_c$ limit
and $A^{ia}$ is the pion axial current. It is found that summing an infinite
class of Feynman diagrams in the leading {\Nc} order corresponds to
solving coupled classical field equations given by the leading order
Lagrangian $L_{ca}+\delta L$.
This produces a baryon source coupled with a classical
meson cloud, with quantum corrections obtained by performing semiclassical
expansion around the classical meson background. This is precisely the
picture described by the chiral bag\cite{rgb83}.

Two aspects of this result are important:
\bitem
\item It is conjectured -- and seems highly plausible --
that there is a line of UV fixed points in the large
{\Nc} renormalization group flow of the parameters of the
Lagrangian\cite{mattis}. The bag radius can be one of those parameters.
If correct, one may formulate CCP in terms of the ``fixed line."
\item The $m_\pi^3$ (or $m_q^{3/2}$) (where $m_\pi$ is the pion mass and $m_q$
the quark mass) non-analytic correction to the baryon mass that is found in the
classical solution is identical to a loop correction in chiral perturbation
theory\cite{manohar}. This makes a direct link between the chiral bag and
chiral perturbation theory ($\chi PT$) to a higher chiral order. We shall
exploit this fact later for nuclear processes and also for Goldstone
boson condensations in dense hadronic matter.
\eitem

\section{Baryon Excitations and Berry Potentials}

\indent

The chiral bag in large $N_c$ limit can be identified with the skyrmion model,
and it describes the baryon ``ground state"  with the semiclassical
configuration in the hedgehog form in flavor $SU(2)$. Collective
rotation built on the hedgehog gives the spectrum with excitations
described by $J=I$ up to $J_{max}=N_c/2$. I would like to discuss
how one can describe other baryonic excitations with $J\neq I$ as well
as excitations involving change of flavor
away from the light-quark (u and d) space. It will
be seen that such baryon excitations can also be described based on
generic symmetry considerations, this time in terms of generalized
Berry potentials. This line of reasoning was developed in
\cite{LNRZ,elafmr} where all other references can be found.

Consider the chiral bag picture. The excitations we wish to consider
can be classified chiefly as follows. One class of excitation
involves quark excitations inside the chiral bag from the lowest
hedgehog state with the grand spin $K=0^+$ where $\vec{K}=\vec{J}+\vec{I}$
to $K\neq 0$ excited quark orbitals. This is an excitation within the $SU(2)$
flavor space. A different class of excitation involves changes of
flavor, corresponding to an excitation from the $K=0^+$ orbital to
a strange (s), charm (c) or bottom (b) orbital. One quark excitation
of either class corresponds to making a ``particle-hole" (p-h) state
in the many-body theory language and such excitations will be coupled on the
bag surface with the corresponding mesons living outside of the
bag, {\it e.g.}, the $K$-meson, $D$-meson, $B$-meson respectively.
For simplicity we will assume that the frequency $\omega_{vib}$
associated with the
p-h excitation (or vibration for short)
is much greater the rotational frequency
$\omega_{rot}$ for the collective rotation of the bag. We shall identify
the vibration as the ``fast" degree of freedom and the rotation as the ``slow"
degree of freedom. The situation then presents a case susceptible to the
Born-Oppenheimer approximation and the idea is to ``integrate out"
the  fast degree of freedom in favor of the slow degree of freedom,
leaving the imprint of the fast degree of freedom in the space of the slow
variable. This is a very generic situation and can be studied with a simple
quantum mechanical system.

\subsection{Analogy with diatomic molecules}
\indent

Following discussion in Ref.\cite{zygelman}, we
consider a generic Hamiltonian describing a system consisting of the
slow (``nuclear") variables $\vec{R} (t)$ representing the
dumb-bell diatom (with $\vec{P}$ as conjugate
momenta) and the fast (``electronic") variables $\vec{r}$
(with $\vec{p}$ as conjugate momenta) coupled through a potential
$V(\vec{R},\vec{r})$
\be
H=\frac{\vec{P}^2}{2M} + \frac{\vec{p}^2}{2m} + V(\vec{R},\vec{r})
\ee
where we use the capitals for the slow variables and lower-case
letters for the fast variables. To describe the symmetry of the system,
let $\vec{N}$ be the unit vector
along the internuclear axis and define the quantum numbers
\be
\Lambda&=&\,\,{\rm eigenvalue\,\,\,of\,\, }\vec{N}\cdot\vec{L} \nonumber \\
\Sigma &=&\,\,{\rm eigenvalue\,\,\,of\,\, }\vec{N}\cdot\vec{S} \nonumber \\
\Omega &=&\,\,{\rm eigenvalue\,\,\,of\,\, }\vec{N}\cdot\vec{J}
=|\Lambda+\Sigma|,
\ee
so $\Lambda,\Sigma,\Omega$ are the projections of the orbital momentum,
spin and total angular momentum of the electron on the molecular axis,
respectively. For simplicity we focus on the simple case of $\Sigma=0$,
$\Lambda=0, \pm 1$. The $\Lambda=0$ state is referred to as $\Sigma$ state
and the $\Lambda=\pm 1$ states are called $\pi$, a degenerate doublet.
We are interested in the property of these triplet states, in particular
in the symmetry associated with their energy splittings.

Upon integrating out the fast electronic degrees of freedom using the
usual adiabatic approximation, we can write the resulting Lagrangian
in the form
\be
L_{nm}^{eff}= \frac{1}{2}M \dot{\vec{R}}(t)^2\delta_{mn}
 + \vec{A}_{mn}[\vec{R}(t)]\cdot
\dot{\vec{R}}(t)-\epsilon_m \delta_{mn}\label{LEFF}
\ee
where
\be
A_{m,n} = i <m|\nabla|n>
\label{generic}
\ee
is a nonabelian Berry potential with the indices $m,n$ labeling
the triplet states.
This can be rewritten in non-matrix form (dropping the trivial
electronic energy $\epsilon$)
\be
{\cal L}=\frac{1}{2}M\dot{\vec{R}}^2 + i\theta^{\dagger}_a(\frac{\partial}
{\partial t}-i\vec{A}^{\alpha}T^{\alpha}_{ab}\cdot\dot{\vec{R}})\theta_b
\label{glagran}
\ee
where we have introduced a Grassmannian variable $\theta_a$ as a trick
to avoid using the matrix form of (\ref{LEFF})
and ${\bf T}^{\alpha}$ is a matrix representation in
the vector space in which the Berry potential lives satisfying the commutation
rule
\be
[{\bf T}^{\alpha},{\bf T}^{\beta}]=i f^{\alpha\beta\gamma}{\bf T}^{\gamma}.
\label{talgeb}
\ee
By a suitable gauge transformation, the Berry potential can be put in
the form that shows its structure as a 't Hooft-Polyakov monopole
\be
\vec{{\bf A}} = (1-\kappa)\frac{\hat{R} \times \vec{I}}{R}
\ee
with the field strength tensor
\be
\vec{{\bf B}}=-(1-\kappa^2)\frac{\hat{R}(\hat{R}\cdot{\bf I})}
{R^2}\label{zgbsf}
\ee
where
\be
\kappa(R)=\frac{1}{\sqrt{2}}|\langle\Sigma|L_x-iL_y|\pi\rangle|.\label{kappa}
\ee
Upon quantizing the theory, one gets the ``hyperfine" spectrum given by
the Hamiltonian
\be
\Delta H = \frac{1}{2M R^2}(\vec{L}_o + (1-\kappa)\vec{I})^2
- \frac{1}{2M R^2}(1-\kappa)^2(\vec{I}\cdot\hat{R})^2\label{hzg1}
\ee
where $L_0$ is the angular momentum of the dumb-bell and $I$ the
angular momentum lodged in the gauge field (inherited from the electronic
sector). The conserved angular momentum is
\be
\vec{L} =  \vec{L}_o + \vec{I},\label{zglf}
\ee
so
\be
\Delta H = \frac{1}{2M R^2}(\vec{L} - \kappa\vec{I})^2
- \frac{1}{2M R^2}(1-\kappa)^2\label{hzg2}
\ee
where  $(\vec{I}\cdot\hat{R})^2 = 1$ has been used. Here $(1-\kappa)$ is
a ``monopole charge" and is clearly not quantized since the value of $\kappa$
is not an integer. Indeed physics lies in the value of $\kappa$ as defined
in (\ref{kappa}).

Let us consider the two extreme cases. The $\pi$ doublet are degenerate
but the $\Sigma$ state is split from the $\pi$ according to the
separation of the dumb-bell $R$. As $R\rightarrow 0$, the energy of the
$\Sigma$ state becomes much higher than the doublet, as a consequence of
which $\kappa\rightarrow 0$ and the spectrum tends to that of the Dirac
monopole with the monopole charge $g=1-\kappa=1$. If on the other hand,
$R\rightarrow \infty$, then the energy of the $\Sigma$ state becomes degenerate
with the doublet and hence $\kappa\rightarrow 1$. In this case, the angular
momentum stored in the gauge field gets decoupled from the system, that is,
the spectrum becomes independent of the angular momentum stored in
the Berry potential.
The consequence is that the electronic rotational symmetry is restored in that
limit. This restoration of the rotational symmetry will have an analog
in the heavy-quark symmetry discussed below.

\subsection{Heavy-quark baryons}
\indent

The above reasoning can be applied almost immediately to the excitation
spectrum of heavy-quark baryons. (One can apply it also to the excitations
of light-quark baryons but the adiabatic approximation is not very good, and
nonadiabatic corrections cannot be ignored. See \cite{LNRZ} on this matter.)
Let $\Phi$ be the ``heavy" mesons
$K$, $D$, $B$ which are the analog to the electronic excitation in the
diatomic case. We then identify the moment of inertia of the rotating soliton
${{\cal I}}$ with $MR^2$, the hyperfine coefficient $c$ with the
``monopole charge" $(1-\kappa)$. Let the angular momentum stored in the
soliton be denoted by $J_{sol}$ and the angular momentum stored in the
gauge field inherited from the heavy mesons be $J_{\Phi}$.
Then the hyperfine spectrum takes the form
\be
\Delta H=\omega_\Phi + \frac{1}{2{{\cal I}}}\left(\vec{J}_R
+ c_\Phi \vec{J}_\Phi\right)^2 +\cdots\label{HF}
\ee
where I have put the ``vibrational" frequency $\omega$ for the heavy meson
$\Phi$ bound in the soliton as described above.
This is a result that follows from the general consideration of
the Berry structure. It was obtained by Callan and Klebanov\cite{CK} for
strange hyperons using a different method.

Given the formula with the moment of inertia ${{\cal I}}$ determined in
the $SU(2)$ soliton sector, we may now take $\omega_\Phi$ and
$c_\Phi$ as parameters for each $\Phi$ and fit the spectra. QCD would
eventually predict those quantities. But for our purpose it is
not essential how one gets them.

The resulting fit is given in Fig.\ref{spectra}
for strange and charm hyperons. One can obtain a similar fit for the
bottom baryons.

\begin{figure}

\vskip 7cm

\caption[spec]{\protect \small
Spectra for strange and charmed hyperons predicted in the model
Eq.(\ref{HF}) compared with the quark model. The fit parameters are
$\omega_K=223$ MeV, $c_K=0.62$, $\omega_D=1418$ MeV, $c_D=0.14$.}
\label{spectra}
\end{figure}

{}From general consideration\cite{MOPR} of large $N_c$ behavior and
heavy-quark symmetry of QCD,
one expects that the constant $c$ should behave
\be
c_\Phi\sim c/m_\Phi
\ee
where $c$ is an $O(N_c^0 m_\Phi^0)$ constant. The fit indeed shows this
with $c\sim 262$ MeV. In the limit $m_\Phi=0$, the Berry potential argument
shows\cite{LNRZ} that $c_\Phi=0$. This is the heavy-quark limit
resembling closely the $R=\infty$ limit of the diatomic molecule.

\section{Chiral Perturbation Theory for Nuclei and Nuclear Matter}
\indent

We have seen that in describing the structure of an {\it elementary}
baryon, the large $N_c$ chiral bag model can be mapped to
chiral perturbation theory in terms of baryons and mesons given as local
fields\cite{manohar}.
We will now take up the same effective chiral Lagrangian and
apply it to many-body systems, {\it i.e.}, nuclei and nuclear matter.

\subsection{Nuclear matter as a Fermi-liquid fixed point}
\indent

There are two classes of physical processes we are interested in.
One is the ground-state property of nuclei and nuclear matter from the
point of view of effective chiral Lagrangians. Given an understanding
of this, we would like to be able to describe the state of matter as
we change the density and /or temperature.

The other is to understand
nuclear force and nuclear response functions in terms of chiral Lagrangians.

The description of these two classes of process would require a
field theory that can describe simultaneously normal nuclear matter
and phase transitions therefrom.
The most relevant ingredient of QCD that is needed here is
spontaneously broken chiral symmetry. For the first, we will be
specifically interested
in chiral $SU(3)\times SU(3)$ symmetry since as we shall see,
strangeness is involved.
In order to address this problem, we need to start from a realistic
effective chiral Lagrangian, obtain a nuclear matter of the right
properties from it and then determine whether a phase change
occurs. For the second, the situation is a bit different and this
``self-consistency" problem can be circumvented in a sense explained below.

At present, we do not have a good  description of nuclear matter
(and nuclei)  starting from
a chiral Lagrangian. There are various suggestions and one promising one is
that nuclear matter arises as a solitonic matter from a chiral effective
action, a sort of chiral liquid\cite{lynn} resembling Landau Fermi liquid.
The hope is that the resulting effective action would look like
Walecka's mean-field model. There is as yet no convincing derivation along
this line. In the work reported here, we will have to assume that we have
a nuclear matter  that comes out of an effective chiral action. Given
such a ground state, we would
like to study fluctuations along various flavor channels and study both
nuclear response functions to slowly varying external fields and possible
instabilities under extreme conditions leading to possible phase transitions.
We are therefore assuming that we can get the properties of normal nuclear
matter (and nuclei) from phenomenology, that is, that nuclear matter is a
Fermi-liquid fixed point\cite{shankar,polchinski}.

In principle, a precise knowledge
of this ground state from a chiral effective Lagrangian at a nonperturbative
QCD level would
allow us to determine the coefficients that appear in the effective Lagrangian
with which to describe fluctuations
around the soliton background -- i.e.,
the Fermi liquid --and with which we could then compute all nuclear response
functions. At present such a derivation does not exist. In a recent paper
by Brown and the author (BR91)\cite{brscaling},
it is assumed that in medium at a matter density $\rho\sim \rho_0$,
the {\it nuclear} effective field theory can be written in terms of
the medium-dependent coupling constants $g^\star$ and masses
of hadrons $m^\star$ while preserving the free-space structure of
a sigma model. This leads to the so-called Brown-Rho scaling. In
\cite{brscaling}, the nonlinear sigma model implemented with trace anomaly
of QCD is used to arrive at the scaling law. The precise way that this
scaling makes sense is elaborated by Adami and Brown\cite{adamibrown} and
in the review (BR94)\cite{newbr94}. I return to this matter below.

\subsection{Chiral counting}
\indent

To describe nuclei and nuclear matter, we need an effective chiral
Lagrangian involving baryons as well
as Goldstone bosons. When baryons are present, \chpt \ is not as firmly
formulated as when they are absent\cite{leutwyler}. The reason is that
the baryon mass $m_B$ is $\sim \Lambda_\chi\sim 1$ GeV, the chiral
symmetry breaking scale. It is more expedient, therefore, to redefine
the baryon field so as to remove the mass from the baryon propagator
\be
B_v=e^{im_B\gamma\cdot v\; v\cdot x} P_+ B
\ee
where $P_+=(1+\gamma\cdot v)/2$ and write the baryon four-momentum
\be
p_\mu=m_B v_\mu +k_\mu
\ee
where $k_\mu$ is the small residual momentum indicating the baryon
being slightly off-shell.
When acted on by a derivative, the baryon field $B_v$ yields
a term of $O(k)$. Chiral perturbation theory in terms of $B_v$ and
Goldstone bosons $(\pi\cdot\lambda/2)$ is known as
``heavy-baryon (HB) \chpt"\cite{JMHBF}. The HB\chpt \ consists of making
chiral expansion in derivatives on Goldstone boson fields,
$\del_M/\Lambda_\chi$,
and on baryon fields, $\del_B/m_B$,  and in the quark mass matrix,
$\kappa {{\cal M}}/\Lambda_\chi^2$. In the meson sector, this is just what
Gasser and Leutwyler did for $\pi\pi$ scattering. In the baryon sector,
consistency with this expansion requires that the chiral counting be made
with $B^\dagger (\cdots)B$, not with $\bar{B} (\cdots)B$. This means that
in medium, it is always the baryon density $\rho (r)$ that comes in and
{\it not} the scalar density $\rho_s (r)$.

Following Weinberg\cite{weinberg}, we organize the chiral expansion
in power $Q^\nu$ where $Q$ is the characteristic energy/momentum scale
we are looking at ($Q<< \Lambda_\chi$) and
\be
\nu=4-N_n-2C+2L +\sum_i \Delta_i\label{counting}
\ee
with the sum over $i$ running over the vertices that appear in the graph
and
\be
\Delta_i=d_i +\frac 12 n_i -2.
\ee
Here $\nu$ gives the power of small momentum (or energy) for a process
involving $N_n$ nucleon lines, %$N_K$ kaon lines,
$L$ number of loops,
$d_i$ number of derivatives (or powers of meson mass) in the $i$th
vertex, $n_i$ number of nucleon lines entering into $i$th vertex and
$C$ is the number of separate connected pieces of the Feynman graph.
In the absence of external gauge fields, chiral invariance requires
that $\Delta_i\geq 0$, so that the leading
power is given by $L=0$, $\nu=4-N_N-2C$. If we are interested in nuclear
responses to external electroweak fields, then $\Delta_i\geq -1$.
\subsection{Nuclear forces from chiral Lagrangians}
\indent

The question as to how much of nuclear forces can be understood starting
from chiral Lagrangians was recently addressed by Weinberg\cite{weinberg}
and by Ord\'o\~nez, Ray and van Kolck\cite{vankolck}. The authors of
\cite{vankolck} studied a chiral Lagrangian consisting of nucleons,
$\Delta$'s and pions applying it to the
nucleon-nucleon potential to the chiral order $\nu=3$
corresponding to $N_n=2$, $C=1$, $L=1$ and $\Delta_i=1$ in (\ref{counting}).
Using a cut-off regularization with a cut-off of order $\Lambda\sim
3.9 {\mbox{fm}}^{-1}$ and fitting the resulting 26 parameters including counter
terms to $I=0$ $np$
and $I=1$ $pp$ phase shifts to $\sim 100$ MeV and to deuteron properties,
they were able to reproduce the global experimental data. The import of
this work is not that it can provide a better potential than what is
currently available in the phenomenological approach but that nuclear
potential can be understood at least at low energy $E\lsim 100$ MeV
from the chiral symmetry point of view.

Effective chiral Lagrangians can also make interesting
statements about nuclear forces in many-body
systems, in particular about many-body forces\cite{weinberg} and exchange
currents\cite{PMR}.

If energy or momentum scale probed $Q$ is much less than the typical
chiral scale $\Lambda_\chi\sim 1$ GeV, then in many-body systems, one
can use static approximation for the pion exchange. In this case, to
the order of chiral expansion that we can actually use at the moment
-- which corresponds to  next-to-next-to-leading ($N^2 L$) order,
three-body forces and currents are exactly canceled with
higher-body forces and currents further suppressed. This justifies
the conventional practice in nuclear physics of ignoring many-body
forces and currents. Of course in higher energy scale at which higher
chiral orders are required, many-body forces and currents
will have to be included. This aspect will be clearly relevant in
the future experiments at CEBAF where multi-GeV energy and momentum
transfers will be involved.

\subsection{Exchange currents}
\indent

Chiral Lagrangians have recently scored an impressive success in
describing exchange vector and axial vector currents at low momentum
transfer. In applying chiral Lagrangians to nuclear response functions,
one has to recognize that while nuclear interactions sample whole range
of distances entering into strong interactions, \chpt \ is applicable
only at sufficiently large distance scales. Thus the most profitable way of
exploiting \chpt \ is to calculate the appropriate amplitude embedded
in the graph sandwiched between initial and final nuclear interactions
that are described by  realistic nuclear potentials. This is the strategy
that has been used since a long time\cite{chemrho}.

\begin{figure}[tbp]
\centerline{\epsfig{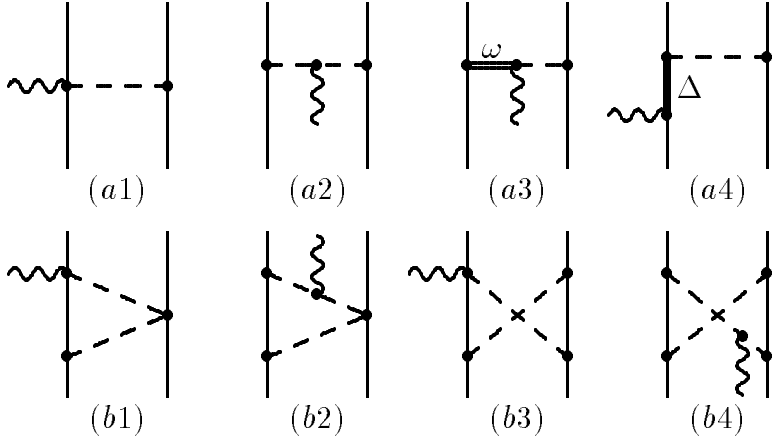}}
\caption{\protect \small
The Feynman graphs which contribute to the two-body vector
current.
The wiggly line is the vector current, the broken line the pion and the
solid line the nucleon.
The ``generalized tree graphs" with the vertices renormalized by loops
are drawn in $(a)$ and the two-pion-exchange graphs in $(b)$.
In Figure $(a)$, the various one loop graphs appearing in
$\pi {\cal V} NN$ vertex are entirely saturated by the resonance-exchange tree
graphs ($(a3)$ and $(a4)$).
Equivalent but topologically different graphs are not drawn. Graphs that
do not contribute in HFF as well as those that have zero measure
in nuclei are also not shown.}
\label{EM}
\end{figure}

Since for slowly
varying external field, three-body, four-body ... currents can be
ignored to the chiral order we will consider, we can focus on
two-nucleon processes. Thus in the counting rule (\ref{counting}),
we will have $N_n=2$, $C=1, 2$ and $\Delta_i\geq -1$. Thus a single-particle
operator will have $\nu=-3$ at the leading order (with $C=2$),
the leading (tree) two-body operator $\nu=-1$ with one-loop corrections
coming in at $\nu=+1$. Here we will calculate to one-loop order, hence
our \chpt \ corresponds to $N^2L$ order.

The most interesting cases to consider are the axial charge operator
$A_0^i$ and the isovector magnetic moment operator $\mu$ coming from the
vector current $\vec{V}^3$. In both cases, the single-particle
operator has an additional $1/m_N$ suppression factor, so its chiral
order is $\nu=-2$ instead of -3. Now to $N^2L$ order in HB\chpt,\ the multitude
of graphs reduce to a handful of them. For instance, for the magnetic moment
operator, there are eight non-vanishing two-body graphs as given
in Fig.\ref{EM}. For the axial-charge operator, there is further reduction
as I will describe below.

\subsubsection{\it Thermal np capture}
\indent

\begin{figure}[tbp]
\centerline{\epsfig{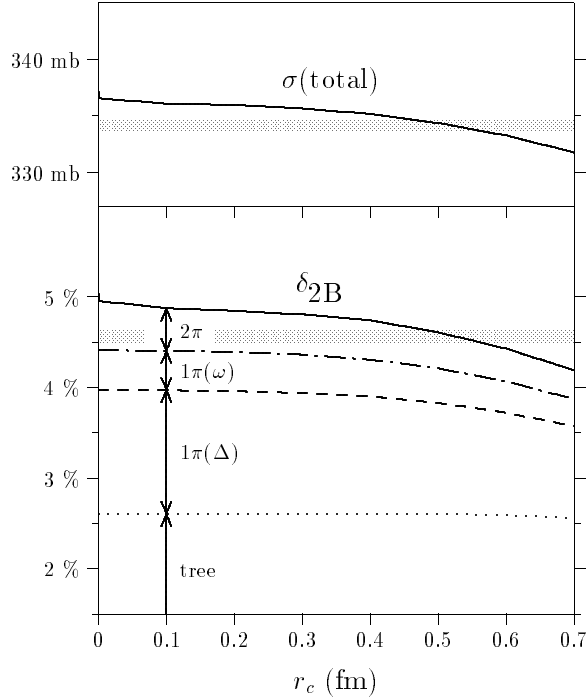}}
\caption{\protect \small
Total capture cross section $\sigma_{\rm cap}$ (top) and $\delta$'s (bottom)
vs. the cut-off $r_c$. The solid line represents the total
contributions and the experimental values are given by the shaded band
indicating the error bar.
The dotted line gives $\delta_{\rm tree}$, the dashed line
$\delta_{\rm tree} + \delta_{1\pi}^\Delta$, the dot-dashed line
$\delta_{\rm tree} + \delta_{1\pi}= \delta_{\rm tree} + \delta_{1\pi}^\Delta
+\delta_{1\pi}^\omega$ and the solid line the total ratio, $\delta_{\rm 2B}$.}
\label{data}
\end{figure}

We first consider the most classic nuclear process\cite{PMR3}
\be
n+p\rightarrow d+\gamma
\ee
which was first explained by Riska and Brown\cite{riskabrown}.
The result of Riska and Brown has recently been reproduced in \chpt \
to $N^2L$ order. The power of \chpt \ is that to $N^2L$ order,
the eight graphs of Fig.\ref{EM} are all there is to calculate.
This clearly goes a considerable distance toward QCD in comparison to
what was achieved in \cite{chemrho}. To this order, the dominant terms are the
``generalized tree" graphs Fig.\ref{EM}(a1)-(a4) with renormalized coupling
constants. The graphs (a1) and (a2) are the leading tree contribution
(called ``tree" ) and the graphs (a3) and (a4) are
$O(Q)$ (or $O(Q^3)$ relative to
the leading term) counter-term contributions that are
saturated by the resonances (called ``1$\pi$" in Fig.\ref{data}).
There are no other operators coming
from the one-loop radiative correction to the vector-N-N-$\pi$
vertex. The remaining two-pion one-loop graphs (b1)-(b4) (called ``2$\pi$")
makes a small
contribution, less than 0.6\% of the single-particle term.

In Fig.\ref{data} is given the result
obtained with the most recent Argonne $v_{18}$ potential of Wiringa, Stoks
and Schiavilla\cite{v18} (which is fit 1787 $pp$ and 2514 $np$
scattering data in the
range 0-350 MeV with a $\chi^2$ of 1.09 and describes the deuteron
properties accurately) compared with the experimental value\cite{cox}.
The impulse operator predicts $\sigma_{imp}=305.6$ mb, about
9.6\% less than the experimental value $\sigma_{exp}=334.2\pm 0.5$ mb.
Using a short-range correlation cutoff $0 < r_c\lsim 0.7$ fm
to screen short-range interactions (which \chpt \ cannot handle),
the theoretical prediction comes out to be
$\sigma_{th}=334\pm 3$ mb in a beautiful agreement with the experiment.
Figure \ref{data}(b) shows how each term contributes to the amplitude
relative to the impulse.

\subsubsection{\it Axial-charge transitions}
\indent

The axial-charge transition
\be
A(J^+)\leftrightarrow B(J^-),\ \ \ \Delta T=1
\ee
in nuclei is known to be enhanced compared with the impulse approximation
prediction. The enhancement can be as much as 100\% in heavy nuclei as
Warburton
has shown\cite{warburton}. This can be understood very simply in terms
of chiral Lagrangians\cite{PMR,ptk94}.

In the case of the axial charge operator, there is further reduction in the
number of diagrams from the vector-current case: Figs.\ref{EM}(a2)
and  (b2) are absent by G-parity, (a4) is suppressed, (a3) has a $\rho$-meson
replacing the $\omega$-meson. There is however one term which is absent in
the case of the magnetic moment operator. This has to do with the fact that
in the graph (a3) it is the $\rho$ meson that contributes. The $\rho$ is
coupled to two pions and it can give a one-loop vertex correction
to the $A_0^i \pi NN$ vertex in contrast to the vertex
$\vec{V}\pi NN$ which receives loop corrections only at two-loop order.

To state the result of the calculation\cite{PMR,ptk94}, we write the
nuclear matrix element of the axial-charge operator as
\be
M^{th}=M_1 + M_2
\ee
with
\be
M_2= M_2^{tree} (1+\delta)
\ee
where the subscript represents the $n=1,2$-body operator and  the ``tree"
corresponds to Fig.\ref{EM}(a1) (with renormalized coupling constants).
It is found that to a good accuracy and almost independently of
mass number\cite{PMR,ptk94},
\be
\delta\lsim 0.1.
\ee
Again as in the electromagnetic case, the tree contribution dominates.
This dominance of the soft-pion process in the cases considered was called
``chiral filter phenomenon." Calculation of the soft-pion term with realistic
wave functions\cite{ptk94} gives a large ratio
\be
M_2^{tree}/M_1\sim 0.6\ -\ 0.8
\ee
enough to explain the experimental value\cite{warburton}
\be
\frac{M^{exp}}{M_1}\sim 1.6\ -\ 2.
\ee
This offers another indication that the pion cloud plays a crucial
role in nuclear processes.

\section{``Swelled" Hadrons in Medium}
\indent

The effective chiral Lagrangian I used so far is a Lagrangian
that results when the degrees of freedom lying
above the chiral scale $\Lambda_\chi$
are eliminated by ``mode integration." As one increases the matter density
or heats the matter,
the scale changes, so we can ask the following question: If a particle moves
in a background with a matter density $\rho$  and/or
temperature $T$, what is the effective Lagrangian applicable in this
background?
One possible approach is to take a theory defined at zero $T$ and zero
$\rho$ and compute what happens as $T$ or $\rho$ increases. This is the
approach
nuclear physicists have been using all along. However now we know that
the major problem with QCD is that the vacuum is very complicated and
we are not sure that by doing the standard approach we are actually describing
the vacuum correctly as $T$ or $\rho$ goes up.
Since the quark condensate $\la \bar{q}q\ra$ is a vacuum
property and it changes as one changes $T$ or $\rho$, it may be more profitable
to change the vacuum appropriate to the given $T$ or $\rho$ and build an
effective theory built on the changed vacuum.
This is the idea of Brown and Rho\cite{brscaling}
in introducing scaled parameters in the effective Lagrangian.

If the quarks are massless, then the QCD Lagrangian is scale-invariant
but quantum mechanically a scale is generated giving rise to the trace anomaly.
In the vacuum, we have in addition to the quark condensate $\la \bar{q}q\ra$
the gluon condensate $\la G_{\mu\nu}G^{\mu\nu}\ra$. We can associate
a scalar field $\chi$ to the $G^2$ field as $G^2\sim \chi^4$ and introduce
the $\chi$ field into the effective Lagrangian to account for the
conformal anomaly of QCD. The $\chi$ field can be decomposed roughly into
two components, one ``smooth" low frequency component and the other
``non-smooth" high-frequency component. The former can be associated
with 2-$\pi$, 4-$\pi$ etc. fluctuations and the latter with a scalar glueball.
For low-energy processes we are interested in, we can integrate out the
high-energy component and work with the low-energy one. In dense matter,
the low-energy component can be identified with a dialton as suggested
by Beane and van Kolck\cite{beane}. Given this identification, one can show
that the BR scaling follows from generic chiral Lagrangians as shown
by Kusaka and Weise\cite{kusaka}. How this can be done in an unambiguous way
is explained in \cite{adamibrown,newbr94}.
The outcome of this operation is that
one can write the same form of the effective Lagrangian as in free space
with the parameters of the theory scaled as
\be
\frac{f_\pi^*}{f_\pi}\approx \frac{m_V^*}{m_V}\approx \frac{m_\sigma^*}
{m_\sigma}\approx\cdots\equiv \Phi (\rho) \label{brscaling1}
\ee
The nucleon effective mass scaled somewhat differently
\be
\frac{m_N^*}{m_N}\approx \sqrt{\frac{g_A^*}{g_A}}\frac{f_\pi^*}{f_\pi}.
\label{brscaling2}
\ee
In these equations the asterisk stands for in-medium quantity.
Now in the skyrmion model, at the mean-field level, $g_A^\star$ does not
scale, so the nucleon will also scale as (\ref{brscaling1}).
It turns out that the pion properties do not scale; the pion mass
remains unchanged in medium at low $T$ and $\rho$.
Thus if one of the ratios in (\ref{brscaling1}) is determined either by
theory or by experiment, then the scaling is completely defined. At
densities up to nuclear matter density, the scaling is roughly
\be
\Phi (\rho)&\approx& 1-a(\rho/\rho_0),\\
a&\approx& 0.15\ -\ 0.2\nonumber
\ee
where $\rho_0$ is the normal nuclear matter density.

Now given the Lagrangian with the scaled parameters, we can go on and do
loop corrections. One of the first things that one finds is that
the $g_A^\star$ gets reduced to $\sim 1$ in nuclear matter from 1.26 in
free space. So one would have to do the whole thing in a consistent way.
However the point is that most of the processes in nuclear physics are
dominated by tree-order diagrams and this means that the effective Lagrangian
with the scaled parameters should be predictive without further corrections.
Indeed this has been what has been found. In a recent paper, Brown, Buballa,
Li and Wambach\cite{BBLW} use this ``BR scaling" to explain simultaneously
the new deep
inelastic muon scattering experiment and Drell-Yan experiments.

A set of rather clear predictions has been made in this
theory\cite{adamibrown,elafmr}.

\section{Kaon Condensation}
\indent

Since I am going to discuss this matter in detail in the Kyoto
Workshop, I shall be rather brief on  this matter.

In a way analogous to describing the baryon mass formula in terms of
the large $N_c$ chiral bag (or skyrmions) and \chpt \
in heavy-baryon formalism,
one can treat kaon condensation in two ways: One in the skyrmion model
and the other in HB\chpt. The two approaches give about the same answer.

\subsection{The Callan-Klebanov Skyrmion on a Hypersphere}
\indent

Callan and Klebanov\cite{CK} suggested in a beautiful paper in 1985 that
in dense medium, the ``effective mass" of the $\overline{K}$ meson
bound in an $SU(2)$ skyrmion could decrease and when it reaches zero,
kaons would condense. This suggestion was examined by Forkel
{\it et al.}\cite{forkel}
by putting the Callan-Klebanov skyrmion on a hypersphere following the
idea of Manton\cite{manton} that the chiral and deconfinement phase
transition(s) could be simulated by putting a single skyrmion on a hypersphere
and by shrinking its radius. Forkel {\it et al.} were interested in the
situation where the kaon mass vanished as would be relevant in heavy-ion
collisions. This was however found
to be impossible except at infinite density or unless the kaon mass
``ran" down as a function of density.  But as shown above, while
the condition that the kaon mass go to zero may be required for condensation
in heavy-ion physics, this is not what is needed in compact-star matter:
It is enough that the mass decrease to the electron chemical potential
$\mu_e$.

A more realistic calculation taking this chemical potential into account
was made recently by Westerberg\cite{westerberg}. For the parameters
of the Skyrme Lagrangian fit to hyperon spectra, the critical density
comes out to be
\be
\rho_c = 0.595\,{\mbox{fm}^{-3}}\simeq 3.5\,\rho_0.
\ee

\subsection{\chpt \ to $N^2L$ order}
\indent

Assuming that nuclear matter at ordinary density is a Fermi-liquid
fixed point, one can look at the fluctuation in the strange flavor direction
by using \chpt. \ This has been recently worked out to $N^2L$ order -- one-loop
in free space and two-loop order in medium -- by Lee {\etal}\cite{LBR}.

There are several issues involved in this calculation.
\bitem
\item The first is to describe
$KN$ scattering at low energy in terms of \chpt. \ To the order considered,
there is no difficulty in doing this.
\item The second is to extend the amplitude to off-shell.
Here on-shell data fit in the first step are not enough to fix all
the parameters of
the Lagrangian. In fact there are two unknown quantities in the counter
terms, one at next-to-leading order and the other at $N^2L$ order.
The first can
be handled by assuming that the counter term is saturated by the decuplet
resonances -- which seems to be a reasonable thing to do. Therefore
we are left with one free parameter. However appearing at subleading
($N^2L$) order, its uncertainty does not affect the calculation for small kaon
frequency which is involved for kaon condensation.
\item  The third issue is
to take into account many-body effects that figure in kaon-nuclear interactions
entering in kaon condensation. Here not only effects on the kaon-nucleon
amplitude embedded in the medium but also intrinsic many-body processes
associated with n-Fermi interactions (for $n\geq 4$)
in the chiral Lagrangian have to be
treated. For this the recent kaonic atom data\cite{kaonicatom} play an
essential role. It turns out that {\it all} the parameters in the chiral
Lagrangian can be fixed by the available kaonic atom data, allowing an almost
parameter-free prediction for the critical density.
\eitem
The result is
\be
2\lsim \frac{\rho_c}{\rho_0}\lsim 4
\ee
in the same range as what is predicted in the skyrmion model.
The lower limit is obtained when Brown-Rho scaling is implemented
in the mean-field (up to $O(Q^2)$) terms. More details can be found
in my Kyoto talk\cite{kyoto}.

\subsection*{Acknowledgments}
I would like to thank G.E. Brown, K. Kubodera, C.-H. Lee, H.K. Lee,
D.-P. Min, M.A. Nowak, T.-S. Park, N.N. Scoccola, S.-J. Sin and
I. Zahed for collaborations and discussions.

\end{document}